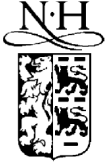

# Performance of large area Micro Pixel Chamber

Tsutomu Nagayoshi[a*], Hidetoshi Kubo[a], Kentaro Miuchi[a], Atsuhiko Ochi[b], Reiko Orito[a], Atsushi Takada[a], Toru Tanimori[a], Masaru Ueno[a],

[a]*Department of Physics, Kyoto University, Kyoto 606-8502, Japan*

[b]*Department of Physics, Kobe University, Kobe 657-8501, Japan*

**Abstract**

A novel gaseous two-dimensional imaging detector "Micro Pixel Chamber (μ-PIC)" has been developed. This detector is based on double sided printed circuit board (PCB). We have developed large area (10cm × 10cm) μ-PICs with 65536 pixel anodes of 400μm pitch on a 100μm thick insulating substrate. Achieved energy resolution was 30% (FWHM) at 5.9keV, and a gas gain of 7000 was obtained with argon ethane (8:2) gas mixture. This gain is high enough to detect minimum ionizing particles with such a small electrode pitch. Although several discharges occurred during 65 hours continuous operation, the detectors have kept stable operation with high gain. The μ-PIC is a useful detector for many applications e.g. X-ray, gamma ray, and charged particle imaging. The micro electrode structure allows us to measure directions of primary electrons due to incident X-rays or gamma rays, which provide a strong method for X-ray polarimetry and gamma-ray imaging. © 2001 Elsevier Science. All rights reserved

## 1. Introduction

Since the invention of the MicroStrip Gas Chamber (MSGC)[1], a large variety of detectors based on the micro processing technology have been developed so far. The most significant requirement for micro pattern gas detectors is the achievement of both stable and high gain operation. Although MSGCs have excellent properties of good position resolution and small space charge effect, they have a serious problem of discharges, which gives critical damage to the detectors.

We proposed a novel gaseous imaging detector having micro electrodes named "Micro Pixel Chamber (μ-PIC)"[2,3]. The μ-PIC has pixel anodes surrounded by cathode rings. A gas avalanche occurs around the anode electrodes due to the strong electric field similar to a wire chamber. In contrast, the electric field is weak at the cathode edges, thus discharge probability is less than MSGCs'. Our μ-PIC is manufactured using printed circuit board (PCB) technology, which allows us to develop a large area (~10cm × 10cm) μ-PIC. Larger ones (up to 30cm × 30cm) will be possible soon. In this paper, we describe the development of large area two-dimensional μ-PICs and this performance.

---

* Corresponding author. Tel.: +81-75-753-3867; fax: +81-75-753-3799; e-mail: nagayosi@cr.scphys.kyoto-u.ac.jp.

## 2. Development

Figure 1 shows a schematic structure of the μ-PIC. This detector is made of a double-sided printed circuit board. Strip anodes and cathodes are orthogonally arranged on both sides of the insulating substrate with a thickness of 100μm. Each cathode strip has holes of 200μm diameter with 400μm pitch. Anode electrodes of ~50μm diameter are formed on the anode strips and pierced at the center of each cathode hole. Thus two-dimensional readout is available using anode and cathode strips. The signal charge from anodes and cathodes are the same size, in contrast to two-dimensional MSGCs, whose pulse height from back electrodes are 20~30% of those from anode electrodes [4]. The thickness of the insulator is similar to the distance between cathode and anode strips. This is the one of features of our μ-PIC compared with the other pixel detectors made of silicon wafer using LSI technology. The electric field due to anode strips is weak at the surface of the insulator. Thus the electric field is not distorted near the substrate by backside electrodes.

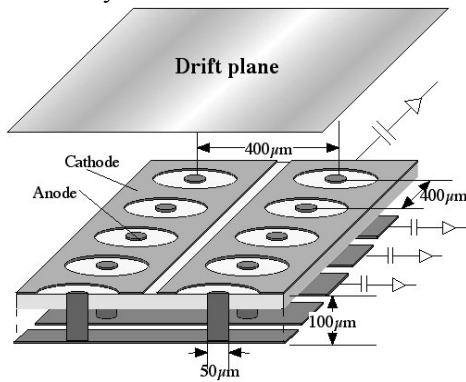

Fig. 1. Schematic structure of the μ-PIC.

Table 1
Summary of our detectors

| Detector | Substrate | Electrodes | Anode height[b] |
|---|---|---|---|
| μ-PIC1 | Ceramics | W, Ni[a], Au[a] | 10μm |
| μ-PIC2 | Polyimide | Cu, Ni[a], Au[a] | -15μm |
| μ-PIC3 | Polyimide | Cu, Ni[a] | -20μm |

[a] Nickel and gold are used for plating.
[b] Distance between the surface of substrate and the top of an anode pixel. Negative value means anode electrodes are below the surface of substrate.

We have developed several types of μ-PICs. Each detector is made of different materials. Additionally, each detector has a different structure of electrode such as the distance between the surface of the substrate and the top of anode electrodes, which are summarized in Table 1. Figure 2 is a microscopic photograph of the μ-PIC1. The μ-PIC has 256 cathodes and 256 anode strips, and covers a 100cm$^2$ detection area. The μ-PIC is mounted on the printed

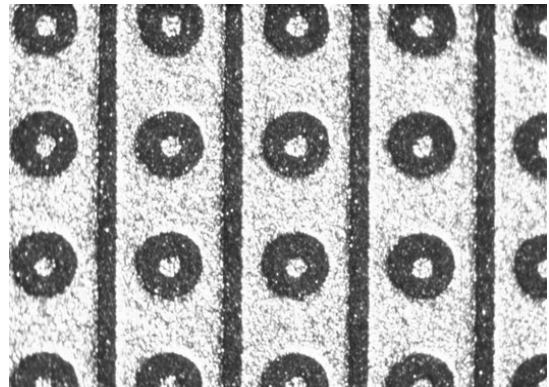

Fig. 2. Microscopic photograph of the μ-PIC viewed from the drift space.

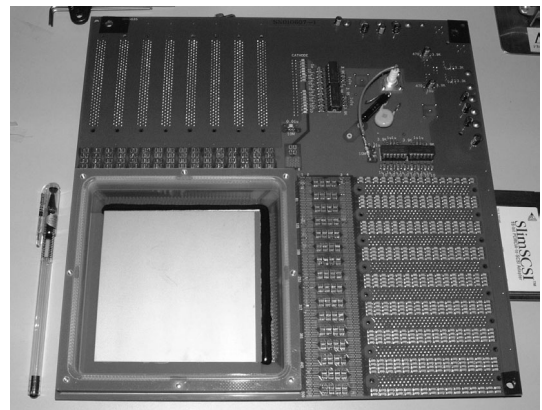

Fig. 3. Photograph of the printed mother board of the μ-PIC.

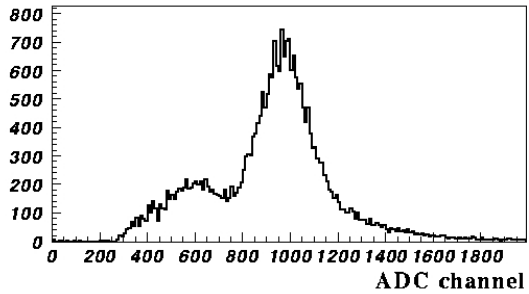

Fig. 4. Cathode pulse height distribution of $^{55}$Fe source measured with the μ-PIC3.

mother board, on which amplifiers are connected to all strips (Fig. 3).

## 3. Performance tests

The detectors were operated in argon ethane (8:2) gas mixture at the atmospheric pressure. The depth of the drift space was 1.2cm and an electric field of 500V/cm was applied. In all operations, we measured the analog summed signal of 16 strips (4096 pixels) by a charged amplifier with a gain of 200mV/pC and a decay time of ~100ns.

Fig. 4 shows an energy spectrum for 5.9keV X-ray of $^{55}$Fe source. An escape peak is clearly seen at 2.7keV as well as the main peak of 5.9keV. Achieved energy resolution was 30% (FWHM). Almost same results were obtained for all μ-PICs. The gas gains were dependent on the position in a detector because of the geometric fluctuation between anode electrodes. Typical gas gain of each detector varies with applied voltage as shown in Fig. 5. In this test, maximum gas gain of 7000 was achieved. Difference of gas gain among the detectors is due to the difference of length of anodes. The insulator between anodes and cathodes degrades gas gain. Therefore higher gas gain was obtained by the detector which has longer anodes.

Stability of μ-PIC depends on the materials of the substrate and electrodes. The electrodes of the ceramic-based detector (μ-PIC1) easily became conductive with resistance of typically several MΩ due to discharges as shown in Fig. 6. It is considered that this conductivity is due to the stuck carbon on the surface of ceramic substrate that is produced from ethane in discharge process, (ceramics have adsorbent and catalytic properties). In contrast, the carbon rarely sticks to the polyimide substrate. Both stable and high gain operation were realized for μ-PIC2 and μ-PIC3 which have polyimide substrate. Therefore, in this case, damage due to discharges is negligible. In the case of μ-PIC2 (gold plated electrode), leakage current of about 100 nA was observed after a large number of discharges, which is probably due to evaporation of gold.

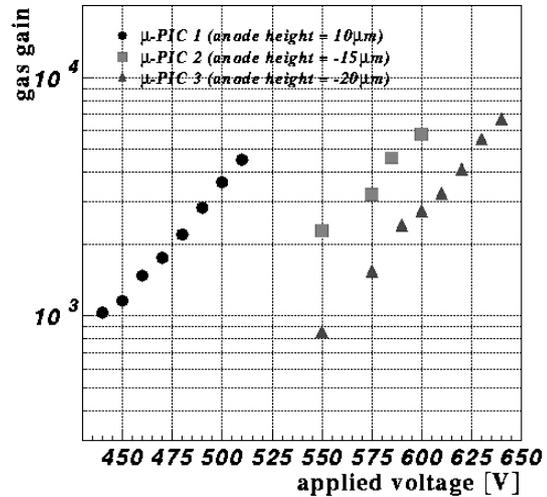

Fig. 5. Gas gains of three types of μ-PICs.

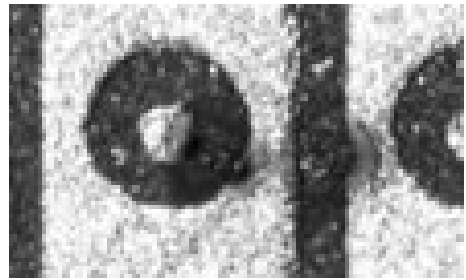

Fig. 6. Microscopic photograph of a damaged electrode of the ceramic-base μ-PIC.

A two-dimensional X-ray image was successfully obtained using a half area of a detector. Figure 7 shows an X-ray transparency image of a key made of iron wrapped in a thin plastic film. The X-ray source is $^{55}$Fe at a distance of 15cm from the drift electrode. An almost uniform image was obtained from about 10000 X-ray events, and structure of millimeter order is apparent. Since the X-ray source was not collimated, the edge of the image is somewhat dispersed. If a collimated beam were available, the position resolution will be similar to that of MSGC [5].

We have confirmed the long time stability via detection of cosmic ray muons. For this purpose, the μ-PIC was used as a Time Projection Chamber (TPC) with 8cm drift region [6]. The detector was operated stably for more than 60 hours with gas gain of 3000. As shown in Fig. 8, there were a few discharges.

## 4. Application of μ-PIC

The micro electrode structure of μ-PIC allows us to observe the track of a primary electron from Compton scattering. This will be a useful tool for gamma-ray imaging of the MeV region. According to computer simulation, about 2 degrees of angular resolution is expected [7]. We are developing a prototype of the gamma-ray imaging detector using μ-PIC, and advanced simulation is now in progress [8].

Detection of X-ray polarization is also possible with the μ-PIC. The pitch of the electrodes is smaller than the range of a photoelectron from an incident X-ray photon. X-ray polarization is obtained using the direction of photoelectrons, because the angular distribution depends on polarization direction [9].

## 5. Summary

A large area μ-PIC has been developed and its performance was measured. A maximum gas gain of 7000, and energy resolution of about 30% (FWHM) at 5.9keV were achieved. Stable operation with enough gain to detect minimum ionizing particles was achieved for polyimide-base μ-PICs. During 65 hours continuous operation, only a few discharges occurred. A Two-dimensional X-ray image was also obtained. The image was almost uniform and millimeter structure of the target was observed, which is similar to the MSGC X-ray imaging system developed previously [5]. Now we are improving the uniformity of the electrodes for larger area μ-PICs.

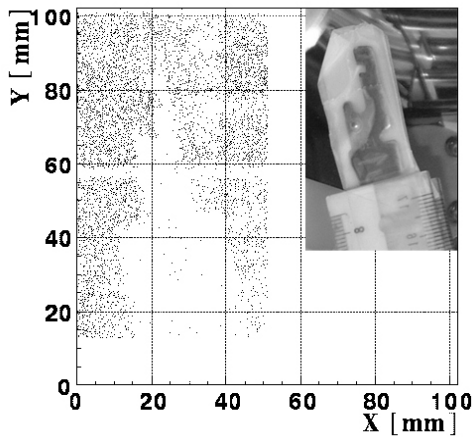

Fig. 7. Two-dimensional image observed by the μ-PIC2. The inset is a key used as the target.

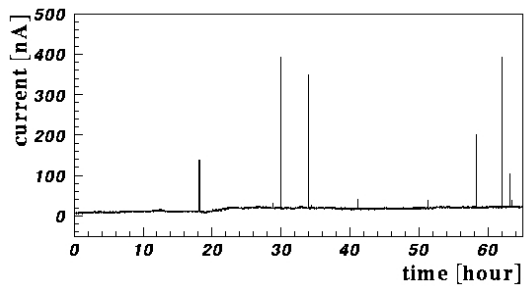

Fig. 8. Leakage current during muon detection test. Spike-shaped peaks mean discharges.

## Acknowledgments

This work is supported by a Grant-in-Aid in Scientific Research of the Japan Ministry of Education, Culture, Science, Sports and Technology, and "Ground Research Announcement for Space Utilization" promoted by Japan Space Forum.

## References

[1] A. Oed, Nucl. Instr. Meth. A263 (1988) 351.
[2] A. Ochi, et al., Nucl. Instr. Meth. A471 (2001) 264.
[3] A. Ochi, et al., Nucl. Instr. Meth. A478 (2002) 196.
[4] T. Tanimori, et al., Nucl. Instr. Meth. A381 (1996) 280.
[5] T. Tanimori, et al., Nucl. Instr. Meth. A436 (1999) 188.
[6] H. Kubo, et al., "Development of the Time Projection Chamber with Micro Pixel Electrodes", these proceedings.
[7] T. Nagayoshi, Proc. 9th EGS4 Users' Meeting, Tsukuba, Japan, 31 Jul.-2 Aug., 2001; KEK-Proc., 2001-22 (2001) 100.
[8] R. Orito, et al., "Development of the MeV Gamma-Ray Imaging Detector with Micro TPC", these proceedings.
[9] A. Ochi, et al., Nucl. Instr. Meth. A392 (1997) 124.